\documentclass[final,5p,times,twocolumn]{elsarticle}

\usepackage{amssymb}
\usepackage{amsmath}

\usepackage[nolist,nohyperlinks]{acronym}
\usepackage{siunitx}
\usepackage{bm}
\usepackage{subcaption}
\usepackage{dblfloatfix}
\usepackage{hyperref}
\usepackage{mathtools}

\usepackage{tikz}
\usepackage{pgfplots}

\pgfplotsset{compat=1.18}

\usepgfplotslibrary{units}
\usepgfplotslibrary{colorbrewer}    
\usetikzlibrary{calc,positioning}
\usetikzlibrary{fit, backgrounds}   
\pgfplotsset{
    unit markings=slash space,
}

\tikzset{
	relative to node/.style={
        shift={(#1.center)},
        x={(#1.east)},
        y={(#1.north)}, 
    }
}

\pgfdeclarelayer{background}
\pgfdeclarelayer{foreground}
\pgfsetlayers{background,main,foreground}

\usepackage{color}
\definecolor{ibilight}{RGB}{193,216,237}
\definecolor{ibidark}{RGB}{0,73,146}	
\definecolor{uke2}{RGB}{170,156,143} 	
\definecolor{uke3}{RGB}{87,87,86}		
\definecolor{ukesec1}{RGB}{255,223,0}	
\definecolor{ukesec2}{RGB}{239,123,5}	
\definecolor{ukesec3}{RGB}{104,195,205}	
\definecolor{ukesec4}{RGB}{138,189,36}	
\definecolor{ukesec5}{RGB}{178,34,41}	
\definecolor{tuhh}{RGB}{45,198,214}     
\definecolor{ibidarkBG}{RGB}{227,229,242}   
\definecolor{uke2BG}{RGB}{233,228,225} 	    
\definecolor{uke3BG}{RGB}{230,231,232}	    
\definecolor{ukesec1BG}{RGB}{255,243,190}   
\definecolor{ukesec2BG}{RGB}{254,232,212}   
\definecolor{ukesec3BG}{RGB}{222,241,241}   
\definecolor{ukesec4BG}{RGB}{233,243,222}   
\definecolor{ukesec5BG}{RGB}{244,230,225}   

\pgfplotsset{
  colormap/vik/.style={%
    /pgfplots/colormap={vik}{%
      rgb=(0.001328,0.069836,0.379529)
      rgb=(0.008947,0.163152,0.438691)
      rgb=(0.010254,0.253516,0.49702)
      rgb=(0.03198,0.349543,0.558906)
      rgb=(0.146607,0.458643,0.631289)
      rgb=(0.317174,0.578997,0.71295)
      rgb=(0.503691,0.69791,0.79398)
      rgb=(0.690457,0.811742,0.871059)
      rgb=(0.868874,0.900206,0.914812)
      rgb=(0.933752,0.858522,0.815421)
      rgb=(0.89269,0.743148,0.657086)
      rgb=(0.844258,0.629296,0.505146)
      rgb=(0.798475,0.52238,0.362835)
      rgb=(0.753316,0.419216,0.22819)
      rgb=(0.684653,0.299264,0.095633)
      rgb=(0.562268,0.168171,0.022523)
      rgb=(0.449521,0.078741,0.025203)
      rgb=(0.350423,6.1e-5,0.030499)
    }
  },
  }

\usepackage{color}
\usepackage{booktabs}
\usepackage{lineno}

\journal{Journal of Magnetic Resonance}

\begin{document}
\begin{acronym}
  \acro{NMR}{nuclear magnetic resonance}
  \acro{MRI}{magnetic resonance imaging}
  \acro{PRESS}{point resolved spectroscopy}
  \acro{RF}{radiofrequency}
  \acro{FID}{free induction decay}
  \acro{CSI}{chemical shift imaging}
  \acro{TSI}{turbo-spectroscopic imaging}
  \acro{EPSI}{echo planar spectroscopic imaging}
  \acro{SNR}{signal-to-noise ratio}
  \acro{mGRE}{multi-gradient echo}
  \acro{FFT}{fast Fourier transform}
  \acro{NFFT}{non-uniform fast Fourier transform}
  \acro{CGNR}{conjugate gradient normal residual}
  \acro{ADMM}{alternating direction method of multipliers}
  \acro{MR}{magnetic resonance}
  \acro{STEAM}{stimulated echo acquisition mode}
  \acro{CI}{confidence interval}
  \acro{IDEAL}{iterative decomposition of water and fat with echo asymmetry and least-squares estimation}
  \acro{VARPRO}{variable projection}
\end{acronym}

\begin{frontmatter}

\title{Rapid quantitative chemical composition mapping\\using model-based MRI reconstruction with field inhomogeneity correction}

\author[ibi,uke]{Artyom Tsanda\corref{cor1}}
\cortext[cor1]{Corresponding author at: Section for Biomedical Imaging, University Medical Center Hamburg-Eppendorf, Lottestraße 55, 22529 Hamburg, Germany.}
\ead{artyom.tsanda@tuhh.de}
\author[ipi]{Stefan Benders}
\author[ipi]{Muhammad Adrian}
\author[ipi]{Alexander Penn}
\author[ibi,uke,imte]{Tobias Knopp}

\affiliation[ibi]{
    organization={Institute for Biomedical Imaging, Hamburg University of Technology},
    addressline={Am Schwarzenberg-Campus 3}, 
    postcode={21073},
    city={Hamburg},
    country={Germany}
}

\affiliation[ipi]{
    organization={Institute of Process Imaging, Hamburg University of Technology},
    addressline={Denickestraße 17}, 
    postcode={21073},
    city={Hamburg}, 
    country={Germany}
}

\affiliation[uke]{
    organization={Section for Biomedical Imaging, University Medical Center Hamburg-Eppendorf},
    addressline={Lottestraße 55}, 
    postcode={22529},
    city={Hamburg}, 
    country={Germany}
}

\affiliation[imte]{
    organization={Fraunhofer Research Institution for Individualized Medical Technology and Engineering IMTE},
    addressline={Mönkhofer Weg 239 a}, 
    postcode={23562},
    city={Lübeck}, 
    country={Germany}
}

\begin{abstract}
    Magnetic resonance spectroscopic imaging methods are particularly attractive for chemical engineering applications, 
    including the monitoring of chemical reactions, where a rapid assessment of spatial variations in chemical composition is required.
    Conventional approaches, such as chemical shift imaging, introduce an additional spectral-encoding dimension, 
    which substantially increases acquisition time. 
    Consequently, fast spatially resolved spectroscopy remains an active research topic.
    This work uses a model-based reconstruction framework that embeds \textit{a priori} spectral knowledge of the involved chemical components into the forward model to accelerate composition mapping. 
    It allows for the reconstruction of molar ratio maps for individual chemical components without acquiring high-resolution spectra.    
    Extending from previous studies, the proposed model accounts for 
    effective frequency offsets caused by main-field inhomogeneity and polarity-dependent readout phase errors, which 
    become more pronounced in systems with larger bores 
    relevant for process engineering.
    Phantom experiments employing a 2D multi-gradient echo sequence demonstrate the ability to determine molar ratios for chemical components with single peaks as well as multiple peaks in their spectra.
    The bias remains in the interval of \SI{-0.056}{mol/mol} to \SI{0.073}{mol/mol} for a \SI{20}{\second} scan, indicating suitability for dynamic processes.
    Finally, a retrospective analysis indicates that acquisition time can be reduced further by applying sparse $k$-space sampling, potentially shortening the scan to \SI{10}{\second} with only minor degradation in quantitative performance.
\end{abstract}




\end{frontmatter}



\section{Introduction}

\Ac{MR} is a valuable measurement technique for chemical engineering applications due to its ability to non-invasively characterize transport and chemical processes within optically opaque systems~\cite{Gladden2017}.
Examples include motion characterization of granular particles and bubble dynamics in gas-solid fluidized beds~\cite{PhysRevE.65.020301, penn_real-time_2017},
quantification of the bubble wake dynamics by measuring velocity fields~\cite{tayler_exploring_2012}, 
temperature distributions in fixed-bed reactors~\cite{serial_temperature_2023, skuntz_experimental_2026}, and 
\textit{operando} characterization of catalytic processes through flow, temperature, and chemical composition measurements~\cite{gladden_mri_2010}, as well as diffusion~\cite{zheng_operando_2023}. 

In particular, \ac{NMR} spectroscopy plays a key role in understanding molecular structure and dynamics. Here, the chemical shift allows signals associated with specific chemical components to be distinguished \cite{arnold_chemical_1951}. 
\Ac{NMR} spectroscopy has broad applications, 
such as monitoring of chemical reactions \cite{blasco_insights_2010, dalitz_process_2012, leutzsch_situ_2019}, 
analysis of protein dynamics \cite{palmer_nmr_2004, angulo_nmr_2024}, 
or studies of lithium-ion batteries \cite{blanc_situ_2013, leifer_nmr_2024}. 
In those applications, the acquired spectrum corresponds to the entire sample placed in the bore.
While spatial variations in chemical composition can sometimes be inferred from other \ac{MR} parameters, such as relaxation times~\cite{evans_magnetic_2004, benders_imaging_2018},
direct spatially resolved spectroscopic information is essential for understanding the complex processes encountered in chemical engineering~\cite{britton_mri_2017}.

In conventional imaging, the frequency encoding is used for one spatial dimension, whereas the phase encoding is employed for the remaining dimensions. The spectral information in the resulting signal is convolved with the spatial encoding and is difficult to recover.
Therefore, conventional approaches for spatially resolved spectroscopy such as \ac{CSI} \cite{brown_nmr_1982} employ phase encoding for all spatial dimensions, yielding a \ac{FID} for each point in $k$-space. Alternatively, a localized single-voxel spectrum can be acquired using one spatially selective \ang{90} excitation pulse followed by two spatially selective \ang{180} refocusing pulses forming an echo. This sequence is called \ac{PRESS} \cite{bottomley_spatial_1987}, 
but there are also other methods with similar strategies such as \ac{STEAM} \cite{frahm1987, frahm1989}.
Since the above-mentioned approaches record a full \ac{FID} with spatial encoding based on phase encoding or slice selection, mapping a volume is time-consuming.

Several methods have been proposed to accelerate the acquisition of spatially resolved spectroscopy~\cite{bogner_accelerated_2021}. 
In \Ac{TSI}~\cite{duyn_fast_1993}, an echo train is used to accelerate acquisition of $k$-space 
by covering multiple points in one acquisition. Due to limited echo time and spin relaxation, 
the spectral resolution of this technique is reduced, but it is used for applications such as 
resolving specific target chemical components relevant for brain metabolism. 
The \ac{EPSI} approach~\cite{mulkern_echo_2001} combines spatial and spectral encoding 
by a rapidly switching frequency encoding gradient, reducing the dimensionality of the problem. 
In this approach, the usage of multiple gradient echoes at different echo times yields 
spectroscopic information via Fourier transformation. 
Therefore, the spectroscopic resolution is limited by the echo spacing and by the number of gradient echoes that can be acquired. Similarly to spin density imaging, spatial encoding for some spectroscopic sequences 
can be optimized by employing different $k$-space trajectories, parallel imaging, and compressed sensing~\cite{bogner_accelerated_2021}. 

For medical imaging, Dixon separated fat and water in images by acquiring at in- and opposed-phase echo times, preserving the spatial resolution of conventional MRI while providing chemical contrast~\cite{dixon_simple_1984}.
Yet, $B_\mathrm{0}$ inhomogeneity interferes with chemical-shift encoding. Three-point acquisitions with phase shifts of $0$, $\pi$, and $2\pi$ resolve this ambiguity and were later implemented within a single fast spin-echo sequence~\cite{glover_multipoint_1991, glover_three-point_1991, hardy_separation_1995}.
To accommodate arbitrary echo times and arbitrary numbers of single-peak components, iterative least-squares fitting of the signal model, hereafter referred to as model-based reconstruction, was proposed~\cite{reeder_multicoil_2004}. The optimization alternates between estimating component intensities and linearly estimating the frequency offset map.
This framework, later named \ac{IDEAL}, was extended to account for relaxation and multi-peak spectra~\cite{reeder_iterative_2005, yu_ideal_2006, yu_multiecho_2008, tsao_hierarchical_2013}.
To improve the robustness of water-fat separation, \ac{VARPRO} decouples offset-map estimation from fat-fraction calculation, estimates the offset map voxel-wise, and enforces its spatial smoothness through regularization~\cite{hernando_joint_2008, hernando_robust_2010}.
Most water-fat methods reconstruct each echo independently in image space, limiting their use for long readouts and sparsely sampled $k$-space trajectories~\cite{daude_comparative_2024}. Doneva \textit{et al.}\ proposed a joint compressed-sensing water-fat model in $k$-space, solved by iterative linearization initialized with fully sampled low-resolution data~\cite{doneva_compressed_2010}.
Nevertheless, water-fat imaging remains largely limited to two components. Multi-peak models require careful calibration because peak weights vary between patients~\cite{yu_ideal_2006}, and jointly estimating $B_\mathrm{0}$ inhomogeneity in the human body remains challenging.
Phantom studies have reported an approximately 2\% bias in fat-fraction quantification~\cite{daude_comparative_2024}. To our knowledge, previous multi-component quantification studies have used water-fat-silicon phantoms containing spatially separated pure components~\cite{tsao_hierarchical_2013}.

Von Harbou \textit{et al.}\cite{harbou} has put the chemical shift imaging into the context of chemical and process engineering.
They formulated a general signal model in $k$-space for multiple chemical components and multi-echo sequences combined with compressed sensing.
To validate the method, two binary mixtures of cyclooctane and 1,4-dioxane were measured 
with a 2D spin echo pulse sequence and undersampled spiral trajectories starting from the center of $k$-space.
With a temporal resolution of \qty{8}{\minute}, the method achieved 
biases of about \SI{0.006}{mol/mol} and \SI{-0.045}{mol/mol} in the estimated dioxane mole fraction for the two mixtures, 
where the larger deviation was attributed to sharp corners and spatial inhomogeneities in the main field $B_\mathrm{0}$.

So far, the chemical shift imaging methods discussed above have assumed static chemical shifts.
However, in reaction monitoring, peak positions may depend on the mixture composition and must therefore be accounted for~\cite{SINGH2019192}.
Chemical shift variations can also occur for some components in polar solvents, for example, for acetone~\cite{monakhova_associationhydrogen_2014}.
If not accounted for, these variations can lead to inaccurate estimates of the chemical composition when the components are distributed heterogeneously in space.

This work further investigates the applicability of the model-based reconstruction for quantitative chemical composition mapping.
It builds upon prior work~\cite{harbou} in two key ways: 
first, by demonstrating that the commonly used Cartesian \ac{mGRE} sequence achieves comparable performance while simplifying implementation on different systems; and second, by overcoming the previous limitation through an extended model that accounts for $B_\mathrm{0}$ inhomogeneities. 
By employing a specific experimental design, where mixtures of interest are submerged into water, the impact of susceptibility artifacts was reduced. Additionally, the introduction of reference water signals allows for direct estimation of the offset map, sufficient for accurate quantitative results unlike prior method for water-fat imaging with compressed sensing~\cite{doneva_compressed_2010}.
Finally, the method is adapted to resolve compositions of components whose chemical shifts depend on mixture composition. The experimental validation assesses quantitative performance in terms of bias and spatial variation across mixing ratios, including chemical components with multi-peak spectra.

\section{Methods}
\subsection{Model}
The signal $s(t) : \mathbb{R}^+ \rightarrow \mathbb{C}$ of a spatially resolved scan with multiple chemical components can be described by the following equation:
\begin{equation}\label{eq:signal}
    s(t) = \int_{V} \int_{\Omega} \rho(\bm{r}, f) e^{-i2\pi f t} e^{-tR(f)} e^{-i\mathbf{k}(t) \cdot \bm{r}} \, df \, d\bm{r} + n(t),
\end{equation}
where $\rho(\bm{r}, f): V \times \Omega \rightarrow \mathbb{C}$ is the spin density of the component at frequency $f$ and location $\bm{r}$, representing the spatially resolved spectrum, 
$V \subset \mathbb{R}^3$ the region of interest, 
$\Omega \subset \mathbb{R}$ the spectral bandwidth, 
$R(f): \Omega \rightarrow \mathbb{R}^+$ the relaxation rate of the component at frequency $f$, 
$\mathbf{k}(t): \mathbb{R}^+ \rightarrow \mathbb{R}^3$ the $k$-space trajectory due to the applied magnetic field gradients for the time period $[0, t]$,
and $n(t): \mathbb{R}^+ \rightarrow \mathbb{C}$ is the noise term. 
Typically, the relaxation rate $R(f)$ corresponds to $1/T_2^*(f)$ for the respective frequency $f$ because it describes the decay of the \ac{FID}.

Estimating fully resolved spectral spin density maps $\rho(\bm{r}, f)$ from the acquired signal $s(t)$ is 
a challenging inverse problem. 
To constrain the solution space, \textit{a priori} knowledge about the chemical composition of the sample can be employed.
In the field of chemical engineering, processes often employ only a finite number of pure chemical components, leading to sparse spectra. 
For this reason, the spatially resolved spectrum in \autoref{eq:signal} can be modeled as a sum of Dirac delta functions, as suggested by von Harbou \textit{et al.}~\cite{harbou}:
\begin{equation}\label{eq:model}
    \rho(\bm{r}, f) = \sum_{k=1}^{M} c_k(\bm{r}) \sum_{j=1}^{L_k} a_{k,j} \delta(f - f_{k,j}),
\end{equation}
where 
$c_k(\bm{r}): V \rightarrow \mathbb{C}$ is the spatial distribution of chemical component $k$ out of the total $M$ components,
$a_{k,j} \in \mathbb{R}^+$ is the amplitude of the $j$-th spectral peak out of $L_k$ peaks of component $k$
corresponding to the frequency $f_{k,j} \in \Omega$. 
Here, the amplitudes $a_{k,j}$ for the component $k$ are calculated analytically using 
\begin{equation}\label{eq:weights}
a_{k,j} = \frac{\rho_k n^{^1H}_{k,j}}{M_k},
\end{equation}
where 
$\rho_k$ is the density of the component $k$ [$\si{\kilogram\per\cubic\meter}$], $n^{^1H}_{k,j}$ is the number of protons contributing to peak $j$, 
and $M_k$ is the molar mass.
Consequently, $a_{k,j}$ has concentration units providing relative weighting of the chemical species, and $c_k(\bm{r})$ remains dimensionless. 
Substituting the model from \autoref{eq:model} into the signal \autoref{eq:signal} and assuming constant relaxation rate $R_k$ for each component $k$ leads to
\begin{equation}\label{eq:signal_and_spectral_model}
    s(t) = \sum_{k=1}^{M} \underbrace{\sum_{j=1}^{L_k} a_{k,j} e^{-i2\pi f_{k,j} t} e^{-t R_k}}_{=\vcentcolon\,\text{CHS}_{k}(t)} \int_{V} c_k(\bm{r}) e^{-i\mathbf{k}(t) \cdot \mathbf{r}} d\mathbf{r} + n(t),
\end{equation}
where $\text{CHS}_{k}(t): \mathbb{R}^+ \rightarrow \mathbb{C}$ denotes the chemical shift term for component $k$ at time $t$. 

Accurate knowledge of the $k$-space trajectory is essential for precise image reconstruction in \ac{MRI}.
Under ideal conditions, the trajectory is derived from the gradient waveforms applied during the imaging sequence.
In practice, system imperfections, such as $B_\mathrm{0}$ inhomogeneity and eddy currents, can lead to frequency offsets deviating calculations from the ideal $k$-space trajectory. 
For example, in conventional imaging, $B_\mathrm{0}$ inhomogeneity can lead to geometric distortions due to errors in spatial encoding. 
In the context of this work, it results in systematic errors in the signal phases, 
leading to inaccurate spectral results.
Therefore, to obtain accurate reconstructions, the signal model has to be extended to account for these effects:
\begin{equation}\label{eq:signal_with_field_inhomogeneity}
    s(t) = \sum_{k=1}^{M} \text{CHS}_{k}(t) \int_{V} c_k(\bm{r}) e^{-i\mathbf{k}(t) \cdot \mathbf{r}} e^{-i\omega_\mathrm{off}(\bm r)t} d\mathbf{r} + n(t),
\end{equation}
where $\omega_\mathrm{off}(\bm r): V \rightarrow \mathbb{R}$ is the spatially varying frequency offset.
Due to the distortions in the Fourier encoding, the integral in \autoref{eq:signal_with_field_inhomogeneity} in a discrete form 
cannot be calculated using the ordinary \ac{FFT} or the more general \ac{NFFT}. 
Instead, one needs tailored algorithms that approximate the off-resonance term 
by a small series consisting of simple Fourier transforms \cite{sutton2003fast}, yielding ${\cal O}(N \log N)$ algorithmic complexity, where $N$ is the number of time points.
Here, an analytical \ac{NFFT}-based method is used to calculate the weighting coefficients for the off-resonance approximation \cite{eggers_field_2007,knopp2008iterative}.
This off-resonance-aware discrete encoding operator can be represented by the matrix $\bm{H} \in \mathbb{C}^{N \times N}$.
The resulting discrete signal $\bm{s} \in \mathbb{C}^{N}$ can be expressed as the superposition of two operators $\textbf{\text{CHS}}_k$ and $\bm{H}$:
\begin{equation}
    \bm{s} = \sum_{k=1}^{M} \textbf{\text{CHS}}_k \cdot \bm{H} \cdot \bm{c}_k + \bm{n},
\end{equation}
where $\textbf{\text{CHS}}_k := \operatorname{diag}(\text{CHS}_k(t_1), \ldots, \text{CHS}_k(t_N)) \in \mathbb{C}^{N \times N}$ is the discrete version of $\text{CHS}_k(t)$,
$\bm{c}_k \in \mathbb{C}^{N}$ is the discrete spatial distribution of the component $k$, and $\bm{n} \in \mathbb{C}^{N}$ is the noise term.

In this work, an \ac{mGRE} sequence is employed with $n_\mathrm{E}$ echoes at echo times $\text{TE}_1, \ldots, \text{TE}_{n_\mathrm{E}}$. 
The signal matrix $\bm{S} \in \mathbb{C}^{n_\mathrm{E} \times N}$ is then composed of the corresponding echo signals.
Under the assumption of instant readouts, elements of the chemical shift term $\textbf{\text{CHS}}_k$ are identical for each respective echo and 
can be combined into a matrix $\textbf{\text{CHS}} \in \mathbb{C}^{n_\mathrm{E} \times M}$
, resulting in the following signal model:
\begin{equation}\label{eq:signal_eq_operator}
    \bm{S} = \textbf{\text{CHS}} \cdot \left(\bm{H} \cdot \bm{C}\right)^T + \bm{N},
\end{equation}
where $\bm{C} \in \mathbb{C}^{N\times M}$ is the spatial distribution of the components, and $\bm{N} \in \mathbb{C}^{n_\mathrm{E}\times N}$ is the noise term.

For non-flyback (bipolar) \ac{mGRE} readouts, the data are acquired during both negative and positive gradient polarities. 
Although the physical B$_0$ inhomogeneity is common to both polarities, the corresponding odd and even echo subsets may contain additional polarity-dependent phase errors caused by eddy currents and other bipolar-gradient imperfections. 
While the resulting phase artifacts can be corrected independently~\cite{yu_phase_2010}, 
this issue can also be addressed within the reconstruction by considering two frequency offsets $\omega_\mathrm{off}^i(\bm r)$, 
where $i \in \{"+", "-"\}$ corresponds to either the positive $(i="+")$ or the negative $(i="-")$ readout direction:
\begin{equation}
    \bm{S}^i = \textbf{\text{CHS}}^i \cdot \left(\bm{H}^i \cdot \bm{C}\right)^T + \bm{N}^i.
\end{equation}
In this case, $\omega_\mathrm{off}^+(\bm r)$ and $\omega_\mathrm{off}^-(\bm r)$ account for both B$_0$ inhomogeneity and polarity-dependent phase discrepancies induced by eddy currents. 
Hereafter, both are referred to as field inhomogeneities or field imperfections.
 
To reconstruct the spatial distribution of the chemical components, the following optimization problem needs to be solved:
\begin{equation}\label{eq:inverse_problem}
    \hat{\bm{C}} = \arg\min_{\bm{C}} \sum_i \left\| \bm{\hat{S}}^i - \textbf{\text{CHS}}^i \cdot \left(\bm{H}^i \cdot \bm{C}\right)^T \right\|_F^2 + \lambda \mathcal{R}(\bm{C}),
\end{equation}
where $\bm{\hat{S}}^i$ is the acquired signal, $\|\cdot\|_F$ denotes the Frobenius norm, $\lambda$ is the regularization parameter, and $\mathcal{R}(\bm{C})$ is a regularization term. 
Unless specified otherwise, Tikhonov regularization $\mathcal{R}(\bm{C}) = \| \bm{C} \|_F^2$ is employed.

As shown by von Harbou \textit{et al.}\cite{harbou}, the method can be further combined with compressed sensing~\cite{1614066, 1580791} and thus extended to sparsely acquired signals $\bm{S}' \in \mathbb{C}^{n_\mathrm{E}\times N'}$, with $N' \leq N$, 
by introducing a binary sampling matrix $\bm{M} \in \{0, 1\}^{N\times N'}$ that selects the acquired $k$-space locations:
\begin{equation}\label{eq:compressed_sensing}
    \hat{\bm{C}} =
    \arg\min_{\bm{C}} \sum_i
    \left\|
        \bm{\hat{S}'}^i -
        \textbf{\text{CHS}}^i \cdot
        \left(\bm{H}^i \cdot \bm{C}\right)^T
        \bm{M}
    \right\|_F^2
    + \lambda \mathcal{R}(\bm{C}).
\end{equation}

Lastly, the signal model can be extended to account for composition-dependent spectra. 
In this case, the chemical shift operator $\textbf{\text{CHS}}^i(\bm{C})$ 
depends on the spatial distribution of the chemical components. For a given $\bm{C}$, 
$\textbf{\text{CHS}}^i(\bm{C})$ maps $\mathbb{C}^{M\times N}$ to $\mathbb{C}^{n_\mathrm{E}\times N}$, 
resulting in a nonlinear reconstruction problem:
\begin{equation}\label{eq:dynamic_spectra}
    \hat{\bm{C}} = \arg\min_{\bm{C}} \sum_i \left\| \bm{\hat{S}}^i - \textbf{\text{CHS}}^i(\bm{C}) \cdot \left(\bm{H}^i \cdot \bm{C}\right)^T \right\|_F^2 + \lambda \mathcal{R}(\bm{C}),
\end{equation}
Unlike the fixed matrix $\textbf{\text{CHS}}^i$ in \autoref{eq:inverse_problem}, 
this operator applies a distinct chemical shift matrix at each voxel according to the local chemical composition. 
Consequently, directly solving \autoref{eq:dynamic_spectra} is computationally demanding. 
In the case when only a single, non-overlapping peak depends on the local composition, the solution can be approximated by solving a series of linear inverse problems. 
This approximation is based on the empirical observation that, at a given voxel, the magnitude of the reconstructed component associated with the selected peak is maximal when its assumed resonance frequency matches that associated with the local chemical composition. 
Accordingly, the linear inverse problem is solved for chemical shift parameterizations that span the expected frequency range. 
For each voxel, the parameterization that maximizes the magnitude of the selected peak is chosen, and the corresponding component estimates form the final reconstruction $\hat{\bm{C}}$.

The resulting reconstructions $\hat{\bm{C}} = \left[\hat{\bm{c}}_1,\ldots, \hat{\bm{c}}_M\right]$, $\hat{\bm{c}}_k \in \mathbb{C}^N$, 
are proportional to the spin concentration of the individual chemical components. The molar ratio $\nu_k$ for each component $k$ can be calculated as follows:
\begin{equation}\label{eq:molar_ratio}
    \nu_k = \frac{\left\| \hat{\bm{c}}_k \right\|_2 / n_{k}^{^1H}}{\sum_{j} \left\| \hat{\bm{c}}_{j} \right\|_2 / n_{j}^{^1H}}
\end{equation}

\subsection{Field inhomogeneity estimation}\label{sec:field_inhomogeneity_estimation}

To account for field imperfections, the proposed model requires the frequency offset maps $\omega_\mathrm{off}^i(\bm r)$ 
as parameters, in addition to the spectral parameters $f_{k,j}$ and $a_{k,j}$ from \autoref{eq:model}.
The $B_\mathrm{0}$ inhomogeneity contributing to the frequency offsets is primarily determined by the magnet and the measurement system.
Additionally, the mismatch in the magnetic susceptibility $\chi$ between different materials can cause the main field to become more inhomogeneous. 
In particular, the magnetic susceptibility mismatch between the sample and the surrounding air induces field disturbances that are most pronounced near the sample boundaries. 
As the inhomogeneities arise from the boundary between different materials, 
the effect is geometrically dependent and challenging to correct \textit{a priori}. 
To mitigate this issue, the experiments described here use a system with substantial length along $B_\mathrm{0}$, 
and the samples are immersed in water. This configuration allows the surrounding water to be used for the estimation of the frequency offsets.

The frequency offset $\omega_\mathrm{off}^i(\bm r)$ is estimated from 
the \ac{mGRE} data by exploiting the linear phase 
evolution across echo times. For each echo time $t_j$, $j \in \{1, \ldots, n_\mathrm{E}^i\}$, out of $n_\mathrm{E}^i$ echoes for one gradient polarity, a spin density image $\rho(\bm r, t_j)$ is reconstructed. 
Water voxels are identified using seeded region growing~\cite{295913},
and phase unwrapping is performed voxel-wise across echo times.
For water voxels, the phase evolves approximately as
\begin{equation}
    \arg\bigl(\rho(\bm r, t_j)\bigr) \approx \phi^i_0(\bm r) + \omega_\mathrm{off}^i(\bm r)\,t_j,
\end{equation}
where $\phi^i_0(\bm r)$ is a time-independent phase offset. 
An ordinary least-squares fit is applied to each water voxel, 
regressing the phase against echo time to estimate the slope $\omega_\mathrm{off}^i(\bm r)$ and intercept $\phi_0^i(\bm r)$.
For the remaining voxels, the frequency offset is obtained from a second-order polynomial fitted to the water-voxel estimates and evaluated over the full field of view.

\subsection{Experimental}
Experiments were conducted using a large-bore vertical \SI{3}{\tesla} \ac{MRI} scanner equipped with a \SI{40}{\cm} diameter radio-frequency coil.
The sample was positioned such that the imaging plane was close to the magnet isocenter to reduce $B_\mathrm{0}$ inhomogeneity across the field of view.

Across all experiments, data were acquired using a 2D \ac{mGRE} sequence.
Unless explicitly stated otherwise, the acquisition used 32 equidistant echoes (TE$_1$ = \SI{1.05}{\milli\second}, echo spacing \SI{0.81}{\milli\second}, TE range \SIrange{1.05}{26.25}{\milli\second}),
repetition time \SI{250}{\milli\second}, flip angle \ang{25}, and one average.
A single \SI{10}{\milli\metre} transverse slice was acquired over a \SI{160}{\milli\metre} \ensuremath{\times} \SI{160}{\milli\metre} field of view using a prescribed 80 \ensuremath{\times} 80 acquisition grid, resulting in a scan duration of \SI{21.5}{\second}.
Phase oversampling with a factor of 2 was applied during acquisition, and the data were cropped in post-processing such that the reconstruction grid matched the prescribed 80 \ensuremath{\times} 80 acquisition grid.

\begin{figure}[t]
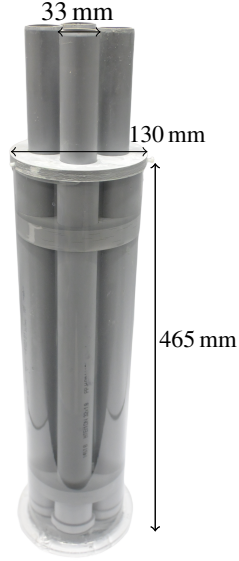

\centering
\include{tikz/sample}
\vspace{-4em}
\caption{
    A sample consisting of a large cylinder made of polymethyl methacrylate
    with cylindrical inserts made of polypropylene. 
    The sample had a height of \qty{465}{\milli\metre}; the diameters were 
    \qty{130}{\milli\metre} for the outer cylinder and 
    \qty{33}{\milli\metre} for the inserts.
    During the experiments, the outer cylinder was filled with water, and the inserts were filled with mixtures of interest.
    The imaging plane was positioned perpendicular to the cylinder axis in the center of the sample.
}\label{fig:sample}
\end{figure}

The sample was a cylinder with a height of \SI{465}{\milli\metre} and a diameter of \SI{130}{\milli\metre}, filled with water (\autoref{fig:sample}). Depending on the experiment, the cylinder contained either one or four polypropylene tubes filled with the mixtures of interest. The tubes had a diameter of \SI{33}{\milli\metre} and were held by three 3D-printed discs to ensure vertical alignment. The transverse imaging slice was placed at the center of the cylinder to reduce susceptibility effects by maximizing the distance to the edges in the $z$ direction.

The proposed model was parametrized with the frequencies and amplitudes of the chemical components, as well as 
the frequency offsets due to field imperfections. The relaxation effect was neglected in the model (see \ref{app:relaxation} for justification).
Unless stated otherwise, all fully sampled data were reconstructed using the \ac{CGNR} solver, with 100 iterations and a regularization parameter of $\lambda = 10^{-4}$.
The reconstruction was implemented in the Julia programming language using \texttt{MRIReco.jl}~\cite{Knopp2021MRIReco} and \texttt{RegularizedLeastSquares.jl}~\cite{hackelberg_regularizedleastsquaresjl_2024} packages.

For quantitative analysis, the bias and spatial variation were considered.
The bias was calculated as the difference between the ground-truth and mean estimated molar ratios within each insert.
The spatial variation was quantified as the standard deviation of the estimated molar ratios within each insert.
For experiments with multiple measurements, the mean bias across measurements was reported together with the corresponding 95\% \acp{CI}, computed using Student's $t$-distribution. The spatial variation was reported as the mean across measurements.
For experiments with a single measurement and multiple inserts, the bias and spatial variation were reported separately for each insert.

In line with the principles of open science, both the source code~\cite{tsanda_code_supplement}
and data~\cite{tsanda_data_supplement} used in this study were made publicly available to ensure reproducibility and support future research.

\subsubsection{Ratio series}

In the first experiment, the performance of the model-based spatially resolved spectroscopy method was investigated for different concentrations of the chemical components. To that end, a series of samples, each consisting of a single cylinder filled with a mixture of water and acetone at a different ratio, was prepared (\autoref{table:model-params-calibration}).

\begin{table}[t]
\centering
\caption{Spectral parameters for acetone-water mixtures used in ratio series experiment. For each ground truth molar ratio $\nu_\text{acetone}$, 
water and acetone frequencies, $f_\text{water}$ and $f_\text{acetone}$, are listed. Small frequency offsets observed for some ratios likely result from imperfect automatic resonance frequency tuning. The error in the ground truth values was estimated by propagating the volumetric uncertainty associated with the graduated cylinder.
}
\label{table:model-params-calibration}
\begin{tabular}{ccc}
\toprule
$\nu_\text{acetone}$ / \si{\mol}/\si{\mol} & $f_\text{water}$ / \si{\hertz} & $f_\text{acetone}$ / \si{\hertz} \\
\midrule
$\num{0.0400}\pm\num{0.0005}$ & \num{0} & \num{-119} \\
$\num{0.125}\pm\num{0.001}$ & \num{40} & \num{-278} \\
$\num{0.250}\pm\num{0.002}$ & \num{40} & \num{-278} \\
$\num{0.500}\pm\num{0.003}$ & \num{0} & \num{-238} \\
$\num{0.600}\pm\num{0.003}$ & \num{0} & \num{-238} \\
$\num{0.800}\pm\num{0.005}$ & \num{0} & \num{-238} \\
$\num{1.0}$ & \num{40} & \num{-198} \\
\bottomrule
\end{tabular}

\end{table}

Due to the small diameter of the cylinder containing the acetone mixtures, $B_\mathrm{0}$ inhomogeneity artifacts within it were negligible. For this reason, \ac{FFT} could be used in \autoref{eq:signal_eq_operator}, and only the frequencies of each component and peak weights needed to be specified as \textit{a priori} knowledge to apply the model-based reconstruction. The amplitudes were computed according to \autoref{eq:weights}, 
namely \num{0.11} for water and \num{0.08} for acetone.
Because the acetone resonance frequency depends on the acetone--water mixing ratio in polar solvents~\cite{monakhova_associationhydrogen_2014}, 
ratio-specific spectral parameterizations estimated from the spectrum of the mean signal within the insert were used. The corresponding values are listed in \autoref{table:model-params-calibration}, where $f_\text{water}$ and $f_\text{acetone}$ denote the single-peak special case of $f_{k,j}$, i.e., $j$ is omitted.

As a baseline, acetone ratios were also estimated from the same \ac{mGRE} data using an \ac{EPSI}-like post-processing approach~\cite{mulkern_echo_2001}.
Since the multi-echo acquisition sampled the complex signal at multiple echo times, 
a discrete Fourier transform along the echo dimension yielded a voxel-wise low-resolution spectrum, 
with bandwidth governed by the echo spacing and resolution by the echo-train duration.
For each voxel, the magnitude of the spectrum was integrated in frequency windows of width \SI{150}{\hertz} centered at 
the water and acetone resonances. The acetone molar ratio $\nu^\text{EPSI}_\text{acetone}$ was computed as 
\begin{equation}
    \nu^\text{EPSI}_\text{acetone} = \frac{I_\text{acetone} / n_\text{acetone}^{^1H}}{I_\text{acetone} / n_\text{acetone}^{^1H} + I_\text{water} / n_\text{water}^{^1H}},
\end{equation}
where $I_\text{acetone}$ and $I_\text{water}$ are the corresponding integrals, and $n_\text{acetone}^{^1H}$ and $n_\text{water}^{^1H}$ denote the number of contributing protons per molecule.

To assess the sensitivity of the model-based reconstruction to inaccuracies in the parameter estimation, an ablation study was performed in which the acetone resonance frequency used in the chemical shift operator was biased by a frequency offset.
Starting from the estimated $f_\text{water}$ and $f_\text{acetone}$, the reconstruction was repeated while keeping all other model parameters fixed and adding a frequency offset $\Delta f$ to $f_\text{acetone}$, with $\Delta f\in\{0,-5,-10,\ldots,-40\}\,\si{\hertz}$.
For the considered mixtures, the resulting acetone molar ratio maps were compared against the nominal mixture ratio within the insert using the voxel-wise absolute errors.

\subsubsection{Validation of frequency offset estimation}

To validate the proposed method for frequency offset estimation described in \autoref{sec:field_inhomogeneity_estimation}, 
an experiment was conducted where both the outer cylinder and four inserts were filled with water. 
In this setup, frequency offsets in the inserts can be estimated directly thereby allowing for a direct comparison 
with the values obtained by extrapolating using polynomial model fitting.
The errors inside the inserts were quantified similar to molar ratios, namely using bias and spatial variation.

\subsubsection{Field inhomogeneity}

To validate the proposed method for field inhomogeneity correction, 
four small cylinders were filled with a mixture of acetone and water at the same ratio $\nu_\text{acetone} = \num{0.250}\pm\num{0.002}$.
Data were reconstructed using the proposed method and the results were compared with the reconstructions obtained without accounting for $B_\mathrm{0}$ inhomogeneity. 
In addition to qualitative comparison of the reconstructions, the distributions of the acetone ratios within the samples were compared.

For comparison with a water--fat separation method, fully constrained \ac{VARPRO}~\cite{ganter_fully-constrained_2026} was applied voxel-wise using \texttt{B0Map.jl}\footnote{https://github.com/cganter/B0Map.jl.git}, 
where acetone was modeled as a single-peak fat component.
Separate fits were performed for the 16 odd and 16 even echoes and restricted to the union of the four insert regions.
The final acetone molar ratio map was obtained by averaging the fat-fraction maps from the two echo subsets.

Furthermore, to demonstrate compatibility with compressed sensing, an ablation study was performed by retrospectively undersampling $k$-space by a factor of \num{2} and reconstructing according to \autoref{eq:compressed_sensing}.
A Cartesian undersampling mask that preserved a small fraction of the central lines (\num{10}\%) and 
randomly selected the remaining $k$-space lines was generated for this purpose (see \ref{app:cs-mask}, 38 lines are retained). 
The undersampled $k$-space was reconstructed using Tikhonov regularization with the \ac{ADMM} solver (\num{30} iterations, $\lambda = 0.01$), 
where the regularization was kept the same for comparability.

\subsubsection{Multi-peak spectra}

To extend the evaluation of the method to components with multi-peak spectra, mixtures of ethanol and acetone were investigated. 
Four mixtures with acetone molar ratios $\nu_\text{acetone} \in \{0.0, 0.300\pm0.002, 0.600\pm0.003, 1.0\}$ were prepared and placed in four inserts in the sample.
For this experiment, the reconstruction model was parameterized with a water resonance at \SI{0}{\hertz}, 
an acetone resonance at \SI{-198}{\hertz}, and three ethanol resonances at \SIlist{-397;-79;159}{\hertz}. Due to limited resolution, J-coupling could be neglected for ethanol. 
The corresponding peak amplitudes were chosen according to \autoref{eq:weights}, 
resulting in amplitudes of \num{0.11} for water, \num{0.08} for acetone, and \num{0.09}, \num{0.04}, and \num{0.03} for the three ethanol peaks, respectively. 

\subsubsection{Composition-dependent spectra}

Because the acetone resonance frequency in polar solvents depends linearly on the acetone molar ratio~\cite{monakhova_associationhydrogen_2014},
mixtures of acetone and water with different ratios were used to evaluate the method for composition-dependent spectra.
Four acetone molar ratios
$\nu_\text{acetone} \in \{0.0577\pm0.0006, 0.140\pm0.001, 0.269\pm0.002, 0.495\pm0.003\}$
were prepared in four inserts.
Reconstructions using fixed spectral parameters were compared with those obtained using the proposed linear approximation.
For the fixed-parameter model, the acetone resonance frequency was set to \SI{-285}{\hertz}.
For the linear approximation, \num{10} uniformly spaced acetone resonance frequencies between \SI{-318}{\hertz} and \SI{-159}{\hertz} were employed.
All other spectral parameters were kept unchanged from the preceding experiments.

\section{Results}

\subsection{Ratio series}

\begin{figure}
\centering
\includegraphics{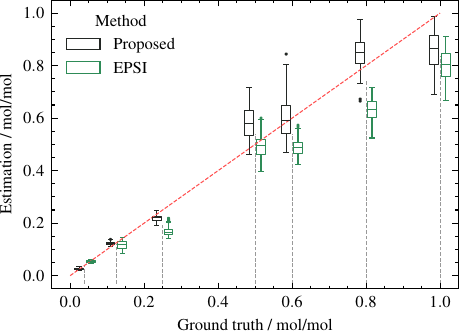}
\caption{
    Ratio series experiment: voxel-wise distributions of the estimated acetone molar ratio within the central insert, plotted against the nominal (ground truth) molar ratio. The proposed model-based reconstruction is compared against an EPSI-like baseline obtained by Fourier transforming the multi-echo signal along the echo-time dimension and integrating the water and acetone resonances. 
    Adjacent boxplots correspond to the same ground truth ratio indicated with the respective vertical dashed lines.
    The diagonal dashed line indicates ideal agreement.
}\label{fig:calibration}
\end{figure}

Voxel-wise distributions of the estimated acetone molar ratios within the insert for different mixtures are shown in \autoref{fig:calibration}.
Good agreement with the ground truth is observed, with a bias of \num{0.008} (95\% \ac{CI}: \num{-0.056} -- \num{0.073}) mol/mol
and a spatial variation of \num{0.043} mol/mol for the acetone molar ratio. 
Accordingly, the positive bias is small compared with the spatial variation between voxels. 
The spatial variation worsens at higher acetone molar ratios.
The baseline \ac{EPSI} method has a bias of \num{0.08} (95\% \ac{CI}: \num{0.00} -- \num{0.16}) mol/mol and
a spatial variation of \num{0.03} mol/mol, 
underestimating the acetone molar ratio and showing lower spatial variation compared to that of the proposed method.

\begin{figure}
\centering
\includegraphics{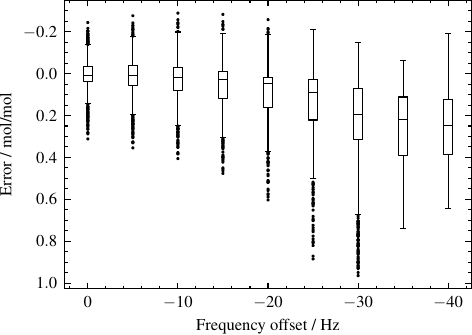}
\caption{
    Sensitivity to parameterization: voxel-wise error of the reconstructed acetone molar ratio within the insert when the acetone resonance frequency in the chemical shift operator is intentionally biased by a frequency offset (all other model parameters fixed).
}\label{fig:sensitivity}
\end{figure}

\autoref{fig:sensitivity} presents the results of the ablation study assessing the sensitivity of the method to bias in the assumed acetone resonance frequency.
The median error increases with increasing magnitude of the frequency offset, and variation increases rapidly with the offset.
For offsets of \SI{-25}{\hertz} and below, the error distribution broadens substantially, spanning a large fraction of the acetone-ratio range. 
This means that the method is sensitive to frequency offsets or bias in \textit{a priori} knowledge.

\subsection{Validation of frequency offset estimation}

\begin{figure*}[!b]
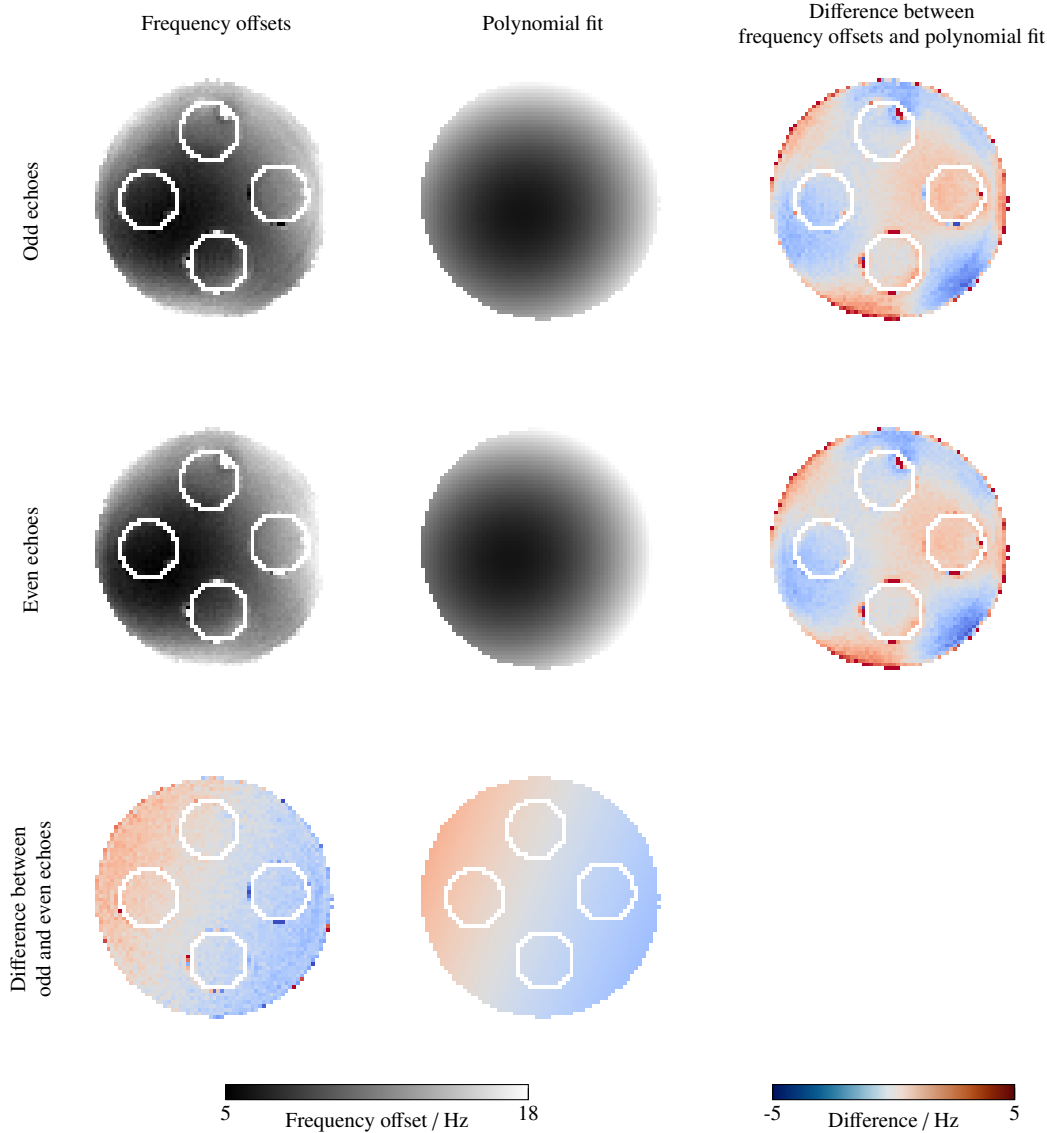

\centering
\include{tikz/B0-map-validation}
\vspace{-0.75cm}
\caption{
    Validation of the proposed method for estimating frequency offsets within the phantom.
    The first column shows frequency offsets estimated from phase evolution in water voxels.
    The second column shows a polynomial fit to the offsets in the outer-cylinder voxels, extrapolated to the inner tubes.
    The third column shows the difference between the measured offsets and the polynomial fit.
    Rows correspond to odd echoes (first row), even echoes (second row), and the difference between odd and even echoes (third row).
}\label{fig:B0-map-validation}
\end{figure*}

\autoref{fig:B0-map-validation} presents the results of the validation experiment for frequency offset estimation.
Polynomial fitting provides a smooth estimate of the frequency offsets across the phantom.
The extrapolated values within the inserts agree with frequency offsets estimated directly from water voxels inside the inserts.
The errors within the inserts are comparable to those across the entire field of view.
The residual error exhibits a systematic pattern across the field of view due to the constraints imposed by the polynomial degree.
The effect from bubbles attached to the insert walls is limited to the immediate vicinity of the bubbles, 
where the largest frequency-offset errors were observed.
For odd echoes, the biases for the inserts are \SIlist{1.2;-0.3;-1.0;0.4}{\hertz}; the spatial variations are \SIlist{1.4;1.5;0.8;1.0}{\hertz}.
For even echoes, the biases for the inserts are \SIlist{1.1;-0.3;-0.9;0.3}{\hertz}; the spatial variations are \SIlist{1.5;1.4;0.6;1.1}{\hertz}.

\subsection{Field inhomogeneity}\label{subsec:fi}

\begin{figure*}
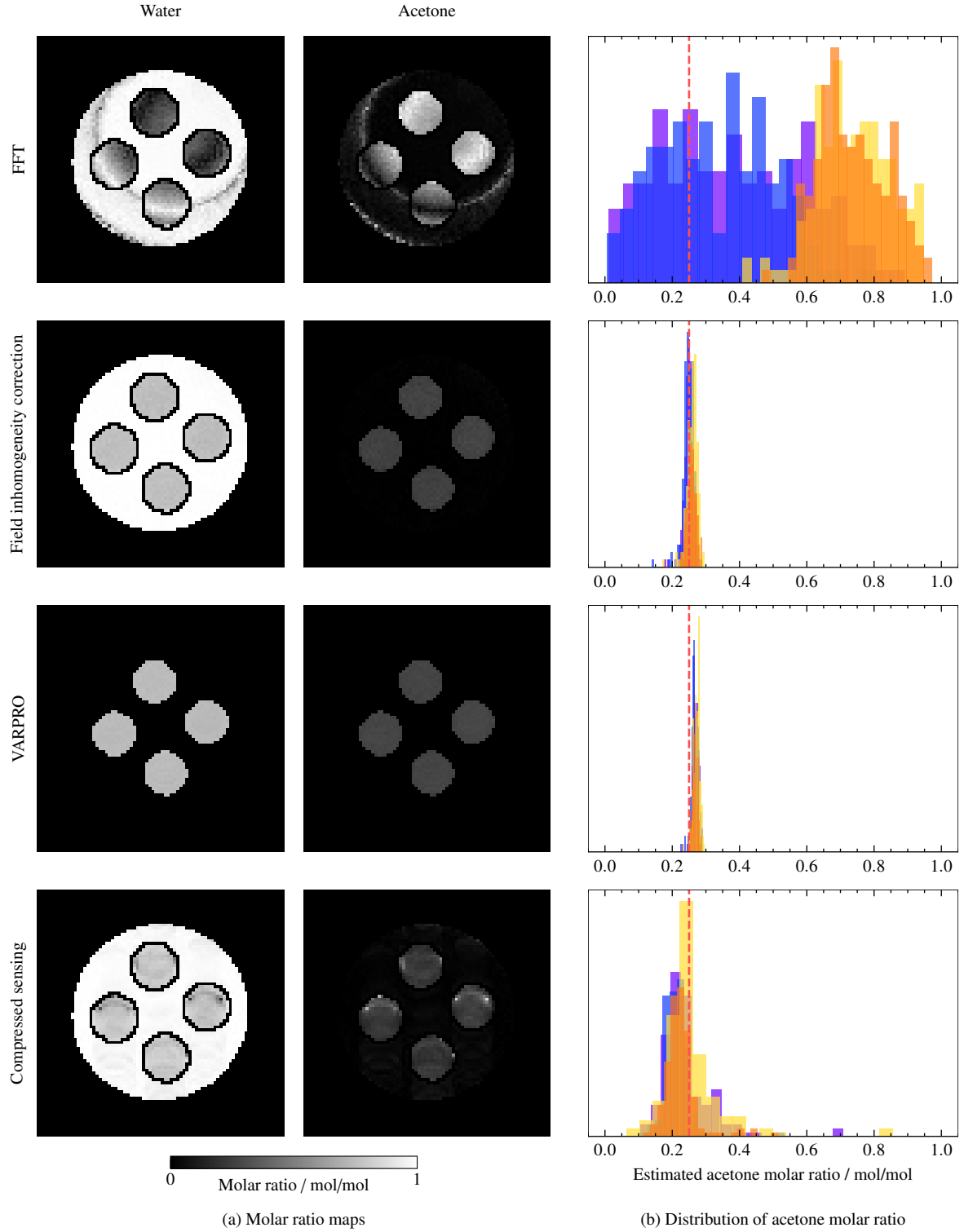

\centering
\include{tikz/field-inhomogeniety_rev1}
\vspace{-0.75cm}
\caption{
    Field inhomogeneity experiment with four identical inserts containing a water--acetone mixture (acetone molar ratio of \num{0.25}) placed at different positions within the water-filled phantom.
    (a)~Water and acetone molar ratio maps reconstructed without field correction (FFT; first row), with field inhomogeneity correction (second row), using fully constrained VARPRO (third row), and from $k$-space retrospectively undersampled by a factor of \num{2} (compressed sensing; fourth row).
    VARPRO was applied only to pixels within the four insert regions.
    (b)~Corresponding distributions of the acetone molar ratio within each insert for the four reconstruction settings. Each color corresponds to one of the four tubes. The dashed red line indicates the ground truth.
}\label{fig:field_inhomogeneity}\phantomsubcaption\label{fig:field_inhomogeneity_a}\phantomsubcaption\label{fig:field_inhomogeneity_b}
\end{figure*}

\begin{figure*}
\centering
\include{tikz/multi-peak-recos}
\vspace{-0.75cm}
\caption{
    Multi-peak spectra experiment with four inserts containing ethanol--acetone mixtures (different acetone fractions) placed at different positions within the water-filled phantom.
    (a) Reconstructed acetone and ethanol molar ratio maps.
    (b) Voxel-wise distributions of the estimated acetone molar ratio within each insert, plotted against nominal (ground truth) molar ratios. The diagonal dashed line indicates ideal agreement.
}\label{fig:spectral_resolution_reconstructions}\phantomsubcaption\label{fig:multi-peak_a}\phantomsubcaption\label{fig:multi-peak_b}
\end{figure*}

\autoref{fig:field_inhomogeneity} shows the results of the field inhomogeneity experiment. 
If uncorrected, field inhomogeneity artifacts cause substantial distortions in the reconstructed acetone and water molar ratio maps (first row in \autoref{fig:field_inhomogeneity_a}). 
The proposed method successfully corrects for the artifacts, providing homogeneous composition maps for both chemical components (second row in \autoref{fig:field_inhomogeneity_a}). 
Quantitatively, for the uncorrected case, 
the biases across the four tubes are \SIlist{-0.13;-0.08;-0.49;-0.49}{mol/mol};
the spatial variations are \SIlist{0.21;0.17;0.11;0.10}{mol/mol}.
For the proposed method, 
the biases are \SIlist{0.0013;0.0023;-0.0092;-0.0077}{mol/mol}; 
the spatial variations are \SIlist{0.0141; 0.0164; 0.0164; 0.0131}{mol/mol}, 
indicating a substantial improvement in both bias and spatial variation.
The disruptions in the images reconstructed with \ac{FFT} result in both biased estimation of the acetone molar ratio and increased spatial variation (see \autoref{fig:field_inhomogeneity_b}). 
With the proposed method, the distributions for each tube align with the ground truth acetone molar ratio of \num{0.25} and have significantly reduced spatial variation.

The fully constrained \Ac{VARPRO} method (third row of \autoref{fig:field_inhomogeneity_a})  successfully corrects the field inhomogeneity artifacts and yields homogeneous molar ratio maps within the inserts. 
Its biases are \SIlist{-0.019;-0.016;-0.022;-0.016}{mol/mol}; spatial variations are \SIlist{0.0087;0.0097;0.0093;0.0078}{mol/mol}.
Although its bias is slightly larger in magnitude than that of the proposed method, the spatial variation is substantially reduced.

Reconstructions from the undersampled $k$-space (last row in \autoref{fig:field_inhomogeneity_a}) match those from the fully sampled acquisition, 
exhibiting aliasing artifacts. These artifacts cause widening of the molar ratio distributions within the inserts. 
Quantitatively, the reconstruction of retrospectively undersampled data yields 
biases of \SIlist{0.008;0.037;-0.007;0.023}{mol/mol}; 
spatial variations of \SIlist{0.080;0.042;0.095;0.053}{mol/mol}.

\subsection{Multi-peak spectra}

\autoref{fig:multi-peak_a} shows the reconstructed acetone and ethanol molar ratio maps and the corresponding acetone molar ratio estimates for the multi-peak spectra experiment.
The two components are clearly separated, and the estimated molar ratios agree with the ground truth values.
Inserts with mixtures exhibit larger deviations than those with pure components, resulting in a noisier texture in the molar ratio maps.
The distributions within each insert (\autoref{fig:multi-peak_b}) follow the expected dependence on the nominal acetone fraction.
Consistent with the molar ratio maps, the insert with acetone molar ratio \num{0.6} shows the largest deviation and the most pronounced outliers.
The biases for the acetone molar ratio across the four tubes are \SIlist{-0.02;0.0;-0.04;0.10}{mol/mol}; 
the spatial variations are \SIlist{0.068;0.031;0.017;0.012}{mol/mol}.

\subsection{Composition-dependent spectra}

\autoref{fig:composition-dependent-spectra_a} shows the reconstructed acetone and water molar ratio maps for the composition-dependent spectra experiment.
The largest discrepancy between the fixed-spectrum and composition-dependent reconstructions occurs in the lower-left insert, which contains the mixture with the highest acetone molar ratio.
At this composition, the acetone resonance frequency is substantially shifted relative to that in the other inserts.
The composition-dependent reconstruction recovers the expected molar ratio in this insert.
This improvement is also evident in the distributions in \autoref{fig:composition-dependent-spectra_b}: the rightmost distribution is close to zero for the fixed-spectrum reconstruction but returns to the expected value for the composition-dependent reconstruction. 
Moreover, the composition-dependent reconstruction recovers the expected monotonic increase in the estimated acetone molar ratio across the four mixing ratios. 
Across all inserts, the fixed-spectrum reconstruction has 
biases of \SIlist{0.04;-0.06;-0.12;0.42}{mol/mol} and spatial variations of \SIlist{0.005;0.015;0.015;0.034}{mol/mol};
while the proposed method has 
biases of \SIlist{-0.01;-0.07;-0.13;-0.04}{mol/mol} and spatial variations of \SIlist{0.005;0.014;0.015;0.042}{mol/mol}.

\begin{figure*}
\centering
\include{tikz/composition-dependent-spectra}
\vspace{-0.75cm}
\caption{
    Experiment with composition-dependent spectra with four inserts containing acetone--water mixtures at different ratios placed within the water-filled phantom. 
    (a) Reconstructed acetone and water molar ratio maps. 
    (b) Voxel-wise distributions of the estimated acetone molar ratio within each insert, plotted against nominal (ground truth) molar ratios. The diagonal dashed line indicates ideal agreement.
}\label{fig:composition-dependent-spectra}\phantomsubcaption\label{fig:composition-dependent-spectra_a}\phantomsubcaption\label{fig:composition-dependent-spectra_b}
\end{figure*}

\section{Discussion}

The results demonstrate the capabilities of model-based chemically resolved MRI for quantitative mixture composition mapping. 
The average bias of \num{0.008} mol/mol in the calibration experiment suggests that the method can reliably quantify mixture compositions 
under near-ideal conditions of two single-peak chemical components with minimal field imperfections. 
The spatial variation worsens at higher acetone molar ratios, which may be due to nonuniform noise propagation from the acquired signal to individual chemical components and the weighting in \autoref{eq:molar_ratio}. 
The field inhomogeneity experiment shows the effectiveness of the proposed method extension, reducing the maximal absolute bias across four inserts from \num{0.49} mol/mol (without correction) to \num{0.0092} mol/mol (with correction).
For the experiment with the ethanol and acetone mixture, the maximum absolute bias remained at \num{0.1} mol/mol, indicating that the method is suitable for components with more complex spectra.
The experiment with composition-dependent spectra showcases a problem relevant for reaction monitoring, where the proposed proof-of-concept method successfully reduces absolute bias for the insert with the highest acetone molar ratio from \num{0.42} mol/mol to \num{0.04} mol/mol.
Across the majority of experiments, the absolute bias remained smaller than the corresponding spatial variation.
While this performance is composition dependent, we expect the method to perform similarly for chemical components having similar distances between peaks.

Compared with other spectroscopic sequences, such as \ac{PRESS} or \ac{CSI}, 
the proposed method substantially reduces the acquisition time (to $\approx\,20$~s), which is comparable to that of conventional MRI sequences.
This acceleration is achieved because the method does not encode the full spectrum at high resolution; instead, it acquires a limited number of echoes
and reconstructs only the frequency components included in the signal model. 
Due to the use of a different sequence, the proposed approach is also faster than the prior art based on the spiral trajectory (\qty{8}{min})~\cite{harbou}.
As shown in \autoref{subsec:fi}, the acquisition time can be further reduced
by employing sparse spatial sampling. 
A twofold retrospective reduction (corresponding to $\approx 10$~s) resulted in a modest increase in bias and more pronounced spatial variation. 
Moreover, employing receive arrays to perform parallel imaging can result in further time savings~\cite{pruessmann1999sense}.
This study kept Tikhonov regularization for comparability. Other regularization strategies, such as Total Variation (TV) or advanced sparse-reconstruction approaches, including deep-learning-based methods~\cite{hammernik_learning_2018}, may achieve higher accuracy.

Reconstructing each echo image independently and applying a Fourier transform along the echo dimension yields a low-resolution spectrum, similar to the \ac{EPSI} method.
By comparison, the proposed model-based reconstruction yields molar ratio estimates (\autoref{fig:calibration}) with lower bias because it fits all echoes jointly within a single forward model.
This joint formulation is particularly advantageous when multiple effects must be modeled simultaneously.
For example, in a sparse \ac{EPSI} acquisition, per-echo reconstructions can accumulate errors across sequential processing steps,
whereas embedding all echoes in a single forward model can mitigate this accumulation.

The method is based on the same physical principles as water-fat imaging, but generalizes these techniques to arbitrary chemical mixtures in chemical and process engineering.
Consequently, for binary mixtures with one or more dominant peaks, established water-fat imaging methods can be directly applied to estimate molar ratios.
This is confirmed by the field-inhomogeneity experiment, in which \ac{VARPRO} corrects for field inhomogeneity and recovers the molar ratios with only a small increase in bias (see \autoref{fig:field_inhomogeneity}).
Yet, joint optimization of frequency offsets is prone to local minima and can make accurate quantification of multicomponent mixtures difficult.
In this work, frequency offset maps were estimated directly from the surrounding water signal used to minimize susceptibility-induced B$_\mathrm{0}$ inhomogeneity.
Although impractical for medical imaging, this approach is feasible in chemical and process engineering applications, where the experimental setup can be adapted.
These applications also introduce challenges not addressed by water-fat imaging, including composition-dependent spectra.
Acetone-water mixtures were used to demonstrate this limitation for which a proof-of-concept solution was proposed.

Model parameterization is the primary limitation of the method. 
As shown in \autoref{fig:sensitivity}, the reconstruction error increases considerably when the assumed frequency parameter for acetone 
is perturbed by only \qty{10}{Hz} (\qty{0.08}{ppm}). 
Consequently, practical applications may require careful calibration of the spectral parameters, for example by recording bulk spectra beforehand.
The ratio-series evaluation was not fully blind because the frequency parameters were estimated from bulk spectra acquired from the same samples used for reconstruction. 
Consequently, the reported performance may be optimistic relative to a strictly independent evaluation. 
The sensitivity analysis in \autoref{fig:sensitivity} complements this evaluation by quantifying the effect of errors in the assumed acetone resonance frequency. 
Because the aim of the method is to recover the spatial distribution of chemical components, bulk spectra provide a valid source of prior information for the reconstruction. 
Moreover, chemical-engineering experiments can include pure components or prepared mixtures for calibration, enabling more accurate spectral parameterization.
Another source of systematic forward-model errors is an unmodeled effective frequency offset arising from 
B$_\mathrm{0}$ inhomogeneity, eddy currents, or other bipolar-gradient imperfections.
In this work, this effect is mitigated by augmenting the forward operator with estimated frequency offset maps employing a polynomial fit.
Although the proposed polynomial fit leaves systematic errors, as shown in \autoref{fig:B0-map-validation}, 
\textcolor{black}{the remaining frequency offset is below \qty{2}{Hz}, which is sufficiently small for accurate reconstruction according to the sensitivity analysis.}
\autoref{fig:field_inhomogeneity} demonstrates that this correction restores the bias to the level achieved in the single-insert experiment, 
where $B_\mathrm{0}$ inhomogeneity was negligible.

Furthermore, the proof-of-concept method for composition-dependent spectra has several limitations.
It is applicable only when a single composition-dependent resonance peak does not overlap with other spectral peaks.
Its voxel-wise maximum-magnitude selection operator is sensitive to noise and systematic model errors, which can bias the estimated resonance frequency and molar ratio.
This sensitivity may contribute to the bias observed in the composition-dependent spectra experiment.
In addition, mitigating susceptibility artifacts imposes geometric constraints on the sample: the air-liquid interface must be sufficiently far
from the imaging plane, which limits applicability for sagittal and coronal slices.

Future work could address these limitations by employing efficient nonlinear optimization techniques and 
by using doping agents to reduce susceptibility differences.
In addition, a broader range of chemical components, including ones with low \ac{SNR} and overlapping spectra, should be evaluated to generate a larger and more diverse dataset, thereby improving the reliability of the proposed method. 
The proposed approach is expected to perform best when spectra are sparse and peaks are well separated, as is the case for some chemical engineering applications.
Accordingly, time-critical dynamic processes, such as chemical reactions, may represent a natural next step for evaluating the method.

\section{Conclusion}
This work demonstrates that model-based \ac{MRI} reconstruction for chemical composition mapping can quantify mixture compositions with low bias and short acquisition times, even in the presence of field imperfections, using \ac{mGRE} sequences. The method is promising for applications requiring fast, quantitative spectroscopic imaging of chemical mixtures.

\section{Declaration of competing interest}
The authors declare that they have no known competing financial
interests or personal relationships that could have appeared to influence
the work reported in this paper.

\section{Acknowledgements}
This project was funded by the Deutsche Forschungsgemeinschaft (DFG, German Research Foundation) – SFB 1615 – 503850735. The authors gratefully acknowledge fruitful discussions with Eric von Harbou, Daniel J. Holland, and Martin Möddel.

\section{CRediT authorship contribution statement}
\noindent\textbf{Artyom Tsanda:} Writing - review \& editing, Writing - original draft, Investigation, Formal analysis, Data curation, Conceptualization. \textbf{Stefan Benders:} Writing - review \& editing, Writing - original draft, Investigation, Formal analysis, Data curation, Conceptualization. \textbf{Muhammad Adrian:} Writing - review \& editing, Investigation, Formal analysis, Data curation, Conceptualization. \textbf{Alexander Penn:} Writing - review \& editing, Funding acquisition, Conceptualization. \textbf{Tobias Knopp:} Writing - review \& editing, Funding acquisition, Conceptualization. 

\section{Data availability}
All datasets are publicly available at \url{https://doi.org/10.15480/882.18630}.

\section{Declaration of generative AI and AI-assisted technologies in the writing process}
This paper was copy-edited with the assistance of GPT-5.4 (OpenAI, San Francisco, CA, USA) in order to improve clarity and readability. The authors have thoroughly checked all the proposed edits and take full responsibility for the content of this manuscript.

\appendix

\section{Compressed-sensing sampling mask}\label{app:cs-mask}

\begin{figure}
    \centering
    \begin{tikzpicture}
	\begin{axis}[
		width=4cm,
		height=4cm,
		scale only axis,
		xmin=0,
		xmax=1,
		ymin=0,
		ymax=161,
		axis x line=none,
		axis y line*=left,
		ytick={1,20,40,60,80,100,120,140,160},
		yticklabels={1,10,20,30,40,50,60,70,80},
		yticklabel style={font=\footnotesize},
		ytick align=outside,
		tick style={black, line width=0.4pt},
		major tick length=2pt,
		enlargelimits=false,
		clip=false,
	]
		\addplot graphics[
			xmin=0,
			xmax=1,
			ymin=0,
			ymax=161,
		] {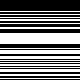};
	\end{axis}
\end{tikzpicture}
    \vspace{-0.75cm}
    \caption{Cartesian $k$-space sampling mask used for the twofold
    retrospective undersampling experiment. Ten percent of the central
    phase-encoding lines were retained, while the remaining samples were
    selected randomly. White corresponds to sampled $k$-space locations.}
    \label{fig:cs-mask}
\end{figure}

\autoref{fig:cs-mask} shows the Cartesian $k$-space sampling mask used for the twofold retrospective undersampling experiment. 
Out of 80 phase-encoding lines (indexed from 1 to 80), the following \num{38} are retained: \numlist{2;5;6;9;10;11;13;15;16;18;20;22;23;24;27;28;29;30;32;37;38;39;40;41;42;43;44;45;46;47;52;53;57;60;61;63;66;69}.

\section{Influence of relaxation}\label{app:relaxation}

This study explicitly ignores relaxation effects in the forward model. 
To motivate this choice, the impact of $T_\mathrm{2}^*$ and $T_\mathrm{1}$ relaxation on the bias of the reconstructed molar ratios is assessed here.

For $T_\mathrm{2}^*$ relaxation, the ratio-series experiment was repeated with the corresponding $T_\mathrm{2}^*$ values incorporated into the forward model.
The $T_2^*$ values were estimated by fitting exponential decay models to the signal and were 
\SI{345}{\milli\second} for water and \SI{324}{\milli\second} for acetone. 
The remaining reconstruction and evaluation steps remained unchanged.
The resulting bias was \num{0.0076} (95\% \ac{CI}: \num{-0.057} -- \num{0.072}) and differed negligibly from that obtained without modeling $T_\mathrm{2}^*$ relaxation.

\begin{figure}[t]
    \centering
    \include{tikz/sequence-variations}
    \vspace{-0.75cm}
    \caption{
        Impact of flip angle (a) and TR (b) variations on the bias of the reconstructed images in field inhomogeneity experiment. 
        Each point represents average bias value in one of four inserts of the phantom.
        Dashed lines and the space between indicates the confidence interval (95\% \ac{CI}: \num{-0.056} -- \num{0.073}) obtained in the ratio-series experiment.
    }
    \label{fig:sequence-variations}
\end{figure}

For $T_\mathrm{1}$ relaxation, the sequence parameters, specifically TR and flip angle, were varied to assess their impact 
on the bias in the field-inhomogeneity experiment. The results are summarized in \autoref{fig:sequence-variations}.
Despite observable variations, all bias values remained within the confidence interval obtained in the ratio-series experiment.
The flip angle of \ang{5} produced the largest deviation in bias relative to the other angles and may be suboptimal for the proposed method.
Lower TR values were associated with greater variance in the estimated bias, presumably because of the lower \ac{SNR}.

For ethanol, accurate estimation of $T_\mathrm{2}^*$ values remained challenging due to multiple peaks and limited number of echoes. The mean signal amplitude showed no visible decay over the echo train and the resulting molar ratio estimates closely matched the expected values.
These observations suggest that relaxation effects are negligible for the proposed study 
but may not generalize to other experimental setups or different compounds where relaxation effects could be more pronounced.

\bibliographystyle{bibstyle} 
\bibliography{refs}

\end{document}